# Adding quantum effects to the semi-classical molecular dynamics simulations

Siyang Yang

Department of Mathematics, Michigan State University, East Lansing, MI 48824, USA

**Abstract:** Simulating the molecular dynamics (MD) using classical or semi-classical trajectories provides important details for the understanding of many chemical reactions, protein folding, drug design, and solvation effects. MD simulations using trajectories have achieved great successes in the computer simulations of various systems, but it is difficult to incorporate quantum effects in a robust way. Therefore, improving quantum wavepacket dynamics and incorporating nonadiabatic transitions and quantum effects into classical and semi-classical molecular dynamics is critical as well as challenging. In this paper, we present a MD scheme in which a new set of equations of motion (EOM) are proposed to effectively propagate nuclear trajectories while conserving quantum mechanical energy which is critical for describing quantum effects like tunneling. The new quantum EOM is tested on a one-state one-dimensional and a two-state two-dimensional model nonadiabatic systems. The global quantum force experienced by each trajectory promotes energy redistribution among the bundle of trajectories, and thus helps the individual trajectory tunnel through the potential barrier higher than the energy of the trajectory itself. Construction of the new quantum force and EOM also provides a better way to treat the issue of back-reaction in mixed quantum-classical (MQC) methods, i.e. self-consistency between quantum degrees of freedom (DOF) and classical DOF.

## I. INTRODUCTION

Direct solution of time-dependent Schrödinger equation (SE) scales exponentially with the number of DOF, so the exact quantum mechanical calculations are currently still limited to the molecular systems involving a small number of atoms. MD utilizing force field[1-13] has been widely implemented for treating condensed phase systems and bio-molecules. Conventional MD is based on two critical approximations: Born-Oppenheimer (BO) separation of electronic and nuclei motion, and classical EOM governing the dynamics of nuclei. For certain chemical reactions involving polarization effects and/or bond breaking and forming,[14-22] dynamics of excited-state is indispensable, and nonadiabatic transitions beyond BO approximation has to be included. Quantum effects like tunneling and zero-point-energy also become critical in many important chemical reactions including proton motion from one water molecule to the next via tunneling.

MQC methods have been widely employed to address the above problems.[23-38] The basic idea lies in the classical-path approximation,[23] which treats the motion of heavy nuclei in a classical manner while evaluating the dynamics of electrons along such classical path of the nuclei. In spite of the huge successes achieved, several specific problems are known to be associated with MQC methods. One major issue of MQC is self-consistency between the quantum and classical DOF,[39] which should both evolve in the way responding to each other correctly. The response of classical DOF to quantum transitions have been accounted for in various MQC methods like Ehrenfest's theorem[25] and surface-hopping (SH) scheme.[32,40] In Ehrenfest's theorem effective force on each trajectory is calculated from the mean-field potential of many coupled potential energy surfaces (PES), while in SH each classical trajectory propagates on a single PES before hopping to another PES happens. In most MQC methods, evolution of quantum DOF under the influence of surrounding classical motion has always been treated in a classical way, i.e. heavy atoms and nuclei propagate according to classical Newtonian EOM to achieve numerical convenience. A truly consistent description should lead to a nonlocal force field for the classical particles, which is lack in MQC methods. Besides, in those methods using classical nuclei trajectories, quantum mechanical effects are accounted for in a quasiclassical sense via a proper initial state sampling. Correspondingly, quantum interferences between different trajectories are not included. Various MQC theories are more suited to specific physical applications. SH was developed for scattering problems with molecular system asymptotically assigned with a single adiabatic electronic state. In a photoreaction, on the other hand, molecular system is prepared in a diabatic electronic state by a laser field.[41,42] Transitions between electronic PESs also play important role in nonadiabatic bound-state relaxation dynamics, e.g. nonradiative photoreactions where a diabatic representation usually provides a better way of understanding the physics dominating the intersecting PESs.

Frozen Gaussian (FG) wavepacket dynamics[43-49] has been designed for describing nonadiabatic effects efficiently and accurately by introducing a unique non-ad hoc mechanism for the breakdown of the BO approximation. Multidimensional frozen Gaussian wavefunctions $\chi_m^I(\mathbf{R})$ are used as nuclear trajectory basis functions (TBF). Classical Hamiltonian EOM are used to propagate the Gaussian centers of all the TBFs. Exact nuclear SE is solved at each time step to calculate the exact complex amplitude of each TBF. FG is different from MQC methods in that all the DOF of the system are treated quantum mechanically on the same footing. Consequently, coherence between basis functions is always conserved, and exact branching ratio can be obtained. There is one issue which remains to be improved: the back-reaction of classical DOF to the dynamics of quantum DOF. In this article, we show how one can improve the EOM governing the nuclei dynamics to incorporate nonlocal force. The new set of quantum-like EOM is designed in the framework of FG, because this method preserves correct branching ratio of trajectories in nonadiabatic transitions as well as removes the "artificial" boundary between the classical and quantum DOF. An important goal in the development of FG was to ensure that it could be applied in the context of *ab initio* molecular dynamics, i.e. *ab-initio* multiple spawning[43] where the electronic SE is solved during the dynamics in order to generate the PESs and inter-state coupling terms. The new EOM accurately conserves quantum mechanical energy of each TBF. This is critical for assigning TBFs with correct tunneling probability. The new quantum EOM not only conserves the quantum energy, but also better conserves the classical energy using newly designed error function. Various classical approximations to the quantum EOM are also discussed in this paper. Particularly, the identical form to the 1$^{st}$ order classical approximation of the current quantum EOM has been used by other methods (e.g. coupled coherent-state technique) to introduce averaged potential which remove the Coulombic singularities.[50]

The current paper is organized as follows. In Section II, a brief introduction to the FG is given. The new EOM is then derived in the framework of FG. Various important classical limits of the new EOM are also discussed. In Section III, the new EOM is tested on a one-state one-dimensional and a two-state two-dimensional model systems. We conclude in Section IV.

**II. THEORY**

*A. Method of FG*

FG utilizes an adaptive, time-dependent basis set of FG, a form pioneered by Heller and coworkers,[51-53] to localize the nuclear SE. Gaussian basis sets have been used in dynamics methods because of their potential in large-scale dynamics calculations.[30,54-68] Unlike the original FG approximation, FG accounts fully for the nonorthogonal nature of the Gaussian basis. The multiconfigurational total wavefunction is written as

$$\psi = \sum_I \chi_I(\mathbf{R};t)\phi_I(\mathbf{r};\mathbf{R}) \tag{1}$$

The subscript $I$ indexes the electronic state, and vectors $\mathbf{r}$ and $\mathbf{R}$ denote the electronic and nuclear coordinates, respectively. The electronic wavefunction $\phi_I(\mathbf{r};\mathbf{R})$ is an eigenfunction of the clamped-nucleus Hamiltonian $\hat{\mathbf{H}}_{el}(\mathbf{r};\mathbf{R})$, obtained by setting the nuclear kinetic energy in the total molecular Hamiltonian to zero. $\chi_I(\mathbf{R};t)$ is the time-dependent nuclear wavefunction associated with the electronic state $I$, and the set of $\chi_I(\mathbf{R};t)$ acts as expansion coefficients in the BO representation.[49] Frozen Gaussians are used to represent the nuclear wavefunction[69,70] for each electronic state,

$$\chi_I(\mathbf{R};t) = \sum_{m=1}^{N_I(t)} c_m^I(t) \chi_m^I\left(\mathbf{R};\bar{\mathbf{R}}_m^I,\bar{\mathbf{P}}_m^I,\gamma_m^I,\alpha_m^I\right) \tag{2}$$

where $N_I(t)$ is the number of nuclear basis functions on electronic state $I$ at time $t$. The value of $N_I(t)$ may change during the propagation, e.g., after spawning. The $c_m^I$ are complex, time-dependent amplitudes paired with each TBF $\chi_m^I$. The full quantum dynamics is described by the bundle of TBFs. Each individual TBF $\chi_m^I$ is written as a multidimensional product of one-dimensional Gaussians $\chi_{m\rho}^I$,

$$\chi_m^I\left(\mathbf{R};\bar{\mathbf{R}}_m^I,\bar{\mathbf{P}}_m^I,\gamma_m^I,\alpha_m^I\right) = e^{i\gamma_m^I(t)} \prod_{\rho=1}^{N_{DOF}} \chi_{m\rho}^I\left(R;\bar{R}_{m\rho}^I,\bar{P}_{m\rho}^I,\alpha_{m\rho}^I\right). \tag{3}$$

where $N_{DOF}$ is the total number of degrees of freedom and each one-dimensional frozen Gaussian is given as

$$\chi_{m\rho}^{I}\left(R; \bar{R}_{m\rho}^{I}, \bar{P}_{m\rho}^{I}, \alpha_{m\rho}^{I}\right) = \left(\frac{2\alpha_{m\rho}^{I}}{\pi}\right)^{1/4} e^{-\alpha_{m\rho}^{I}\left(R-\bar{R}_{m\rho}^{I}\right)^{2} + i\bar{P}_{m\rho}^{I}\left(R-\bar{R}_{m\rho}^{I}\right)} \quad (4)$$

The FGs are parameterized by centroid position $\bar{R}_{m\rho}^{I}$, momentum $\bar{P}_{m\rho}^{I}$, width $\alpha_{m\rho}^{I}$, and phase $\gamma_{m}^{I}$. Numerical considerations encourage the use of a fixed, time-independent width parameter $\alpha_{m\rho}^{I}$. In general, simulation results are insensitive to the particular values chosen for $\alpha_{m\rho}^{I}$, so long as they fall within a fairly broad range.[71,72] In simulation, the value of $\alpha_{m\rho}^{I}$ is only determined by $\rho$, which indexes DOF, while is constant for all TBFs on various PESs, therefore one simply denotes it by $\alpha_{\rho}$. The position and momentum centers $\bar{\mathbf{R}}_{m}^{I}$ and $\bar{\mathbf{P}}_{m}^{I}$ of the TBFs evolve according to classical mechanics on the *I*-th electronic state

$$\begin{aligned} \frac{\partial \bar{R}_{j}}{\partial t} &= \frac{\bar{P}_{j}}{m} \\ \frac{\partial \bar{P}_{j}}{\partial t} &= -\left.\frac{\partial V(R_{j})}{\partial R_{j}}\right|_{R_{j}=\bar{R}_{j}} \end{aligned} \quad (5)$$

Though the center of Gaussian for each TBF propagates classically, full quantum mechanics is recovered or retained through the EOM governing the complex amplitude,

$$\frac{d\mathbf{C}^{I}}{dt} = -i\left(\mathbf{S}^{II}\right)^{-1}\left\{\left(\mathbf{H}^{II} - i\dot{\mathbf{S}}^{II}\right)\mathbf{C}^{I} + \sum_{J \neq I}\mathbf{H}^{IJ}\mathbf{C}^{J}\right\} \quad (6)$$

where $\mathbf{C}^{I} = \{c_{m}^{I}; m = 1, 2, ..., N\}$ is composed of the complex coefficients $c_{m}^{I}$ for each TBF, and the tensors **H** and **S** are defined by

$$H_{ij}^{IJ} \equiv \left\langle \chi_i^I(\mathbf{R};t) \middle| \left\langle \phi_I(\mathbf{r};\mathbf{R}) \middle| \hat{\mathbf{H}} \middle| \phi_J(\mathbf{r};\mathbf{R}) \right\rangle_{\mathbf{r}} \middle| \chi_j^J(\mathbf{R};t) \right\rangle_{\mathbf{R}}$$

$$S_{ij}^{IJ} \equiv \left\langle \chi_i^I(\mathbf{R};t) \middle| \left\langle \phi_I(\mathbf{r};\mathbf{R}) \middle| \hat{\mathbf{1}} \middle| \phi_J(\mathbf{r};\mathbf{R}) \right\rangle_{\mathbf{r}} \middle| \chi_j^J(\mathbf{R};t) \right\rangle_{\mathbf{R}}$$

(7)

Note that, for convenience, $\mathbf{H}^{IJ}$ is also called matrix with matrix elements $H_{ij}^{IJ}$ defined above. The overlap matrix accounts for the nonorthogonal character of the nuclear basis functions. The right-acting time derivative of $S_{ij}^{IJ}$ is defined by

$$\dot{S}_{mn}^{IJ} = \delta_{IJ} \left\langle \chi_m^I(\mathbf{R};t) \middle| \frac{d}{dt} \chi_n^J(\mathbf{R};t) \right\rangle_{\mathbf{R}}$$

(8)

which expresses how the non-orthogonality of the nuclear basis functions changes in time.

In the limit of a complete nuclear basis set on all electronic states, the prescription given so far would lead to a numerically exact solution of the SE. In practice, methods that employ the FG, e.g. the fully multiple spawning (FMS) method[44,46], usually start with a small basis set. Since electronic transitions are usually marked by strong nonadiabatic coupling, FG allows TBFs on one electronic state to spawn new TBFs on another electronic state only when they are in a region with significant nonadiabatic coupling. This spawning algorithm adaptively increases the basis set while maintaining an accurate description of the evolving wavefunction.

*B. New equations of motion conserving quantum mechanical energy*

Energy conservation is an important issue for dynamics. In FMS, each TBF evolves according to classical EOM, and the energy of the parent and that of spawned child TBF are constrained to be identical, so the classical energy of each individual TBF is conserved, in agreement with standard MD and momentum-jump in SH.[39] In many-dimensional systems, any finite number of initial basis functions will be negligibly coupled in the long time limit due to exponential divergence in phase space.[50] One can safely assume that the trajectory ensemble will behave in the long time limit as an incoherent superposition of independent basis functions. The energy computed as the expectation value of the Hamiltonian will be essentially a population-weighted average of the classical energies of the TBFs.

As a result of conservation of classical energy, quantum mechanical energy is not conserved in FMS. This feature of FMS actually improves the numerical efficiency without causing serious problem in general, e.g. correct branching ratio can still be obtained. Besides, the fluctuation of quantum mechanical energy decreases as the FMS basis set increases. However, strict conservation of quantum mechanical energy is critical if one wants to describe tunneling using bundle of TBFs, because the tunneling probability depends sensitively on the energy of both the potential barrier and each individual TBF.

For most *ab initio* calculations, numerical convenience obtained via classical equations of motion conflicts sharply with the requirement of strict conservation of quantum mechanical energy, since the latter requires calculation of nonlocal quantum force which is usually computationally expensive. In principle, any kind of MD conducted in phase space inevitably suffers from the loss of information when reduced from the infinite-dimensional Hilbert space to classical phase space. However, this principle does not forbid the possibility of finding EOM conserving the true quantum mechanical energy only, which is the goal of current paper. Time-dependent variational principle (TDVP)[73] provides one promising and powerful tool for the design of new quantum mechanical EOM, but we prefer the scheme that can be related to certain classical picture in one limit. The existence of asymptotic limit of classical mechanics guarantees the intuitional physical picture and our understanding about the dynamics.

We now develop a new set of EOM that conserves quantum mechanical energy while retaining classical limit. The derivation is carried out in the framework of FMS, but we believe the underlying principle applies to general category including various MQC and semiclassical (SC) methods. Conservation of quantum mechanical energy reads $\frac{d}{dt}\langle\psi(\mathbf{r},\mathbf{R})|\mathbf{H}|\psi(\mathbf{r},\mathbf{R})\rangle = 0$ where $E_{QM} = \langle\psi(\mathbf{r},\mathbf{R})|\mathbf{H}|\psi(\mathbf{r},\mathbf{R})\rangle$ is the expectation value of Hamiltonian projected onto the total wavefunction $\psi$. Making use of (6), one can show that conservation of quantum energy requires the following equation to be satisfied,

$$\begin{aligned}
0 &= \frac{d}{dt}\sum_{I,J}\mathbf{C}^{I\dagger}\cdot\mathbf{H}^{IJ}\cdot\mathbf{C}^J = \sum_{I,J}\left[i\mathbf{C}^{I\dagger}\left(\mathbf{H}^{II}+i\dot{\mathbf{S}}^{II}\right)\mathbf{S}^{II^{-1}} + i\sum_{K\neq I}\mathbf{C}^{K\dagger}\mathbf{H}^{KI}\mathbf{S}^{II^{-1}}\right]\mathbf{H}^{IJ}\mathbf{C}^J \\
&+ \sum_{I,J}\mathbf{C}^{I\dagger}\mathbf{H}^{IJ}\left[-i\mathbf{S}^{JJ^{-1}}\left(\mathbf{H}^{JJ}-i\dot{\mathbf{S}}^{JJ}\right)\mathbf{C}^J - i\mathbf{S}^{JJ^{-1}}\sum_{K\neq J}\mathbf{H}^{JK}\mathbf{C}^K\right] + \mathbf{C}^{I\dagger}\dot{\mathbf{H}}^{IJ}\mathbf{C}^J \\
&= \sum_{I,J}\mathbf{C}^{I\dagger}\left(\dot{\mathbf{H}}^{IJ}-\dot{\mathbf{S}}^{II}\mathbf{S}^{II^{-1}}\mathbf{H}^{IJ}-\mathbf{H}^{IJ}\mathbf{S}^{JJ^{-1}}\dot{\mathbf{S}}^{JJ}\right)\mathbf{C}^J \\
&+ i\sum_{I,J,K}\left(\mathbf{C}^{K\dagger}\mathbf{H}^{KJ}\mathbf{S}^{JJ^{-1}}\mathbf{H}^{JI}\mathbf{C}^I - \mathbf{C}^{I\dagger}\mathbf{H}^{IJ}\mathbf{S}^{JJ^{-1}}\mathbf{H}^{JK}\mathbf{C}^K\right) \quad (9)
\end{aligned}$$

where matrix $\mathbf{H}^{IJ}$ is defined by (7), and $\dot{\mathbf{H}}^{IJ} \equiv \frac{d\mathbf{H}^{IJ}}{dt}$. The summation is over all the PESs. For simplicity, we consider the case of single PES in the following derivation. Eq. (9) can be simplified as

$$0 = \mathbf{C}^{\dagger}\cdot\left(\dot{\mathbf{H}}-\dot{\mathbf{S}}\mathbf{S}^{-1}\mathbf{H}-\mathbf{H}\mathbf{S}^{-1}\dot{\mathbf{S}}\right)\cdot\mathbf{C} \quad (10)$$

where the superscripts of $\mathbf{H}$ and $\mathbf{S}$ have been dropped since only one PES exists. The matrix elements can be evaluated as follows,

$$\dot{H}_{ij} = H_{ij}^{10}\left(2\alpha\dot{\bar{R}}_i - i\dot{\bar{P}}_i\right) + H_{ij}^{01}\left(2\alpha\dot{\bar{R}}_j + i\dot{\bar{P}}_j\right) + H_{ij}\left(i\dot{\bar{R}}_i\bar{P}_i - i\dot{\bar{R}}_j\bar{P}_j - i\dot{\gamma}_i + i\dot{\gamma}_j\right)$$

$$\dot{S}_{ij} = \left\langle \chi_i \left| \frac{d}{dt} \chi_j \right\rangle = S_{ij}\left(-i\dot{\bar{R}}_j\bar{P}_j + i\dot{\gamma}_j\right) + S_{ij}^{01}\left(2\alpha\dot{\bar{R}}_j + i\dot{\bar{P}}_j\right)$$

$$\dot{\tilde{S}}_{ij} = \left\langle \frac{d}{dt}\chi_i \middle| \chi_j \right\rangle = S_{ij}\left(i\dot{\bar{R}}_i\bar{P}_i - i\dot{\gamma}_i\right) + S_{ij}^{10}\left(2\alpha\dot{\bar{R}}_i - i\dot{\bar{P}}_i\right) \tag{11}$$

where, for simplicity, one-dimension position and momentum vectors are assumed in the above derivation. Generalization to multi-dimensional case is straight-forward as shown in the next section. The one-dimensional TBF $\chi_i(R;\bar{R}_i,\bar{P}_i,\gamma_i,\alpha)$ is defined by (3). Therefore, one can evaluate the matrix elements

$$\begin{aligned}
&\left(\dot{\mathbf{H}} - \dot{\tilde{\mathbf{S}}}\mathbf{S}^{-1}\mathbf{H} - \mathbf{H}\mathbf{S}^{-1}\dot{\mathbf{S}}\right)_{ij} \\
&= \left(2\alpha\dot{\bar{R}}_i - i\dot{\bar{P}}_i\right)\left(H^{10} - S^{10}S^{-1}H\right)_{ij} + \left(2\alpha\dot{\bar{R}}_j + i\dot{\bar{P}}_j\right)\left(H^{01} - HS^{-1}S^{01}\right)_{ij} \\
&\quad + \left(i\dot{\bar{R}}_i\bar{P}_i - i\dot{\gamma}_i\right)\left(H - SS^{-1}H\right)_{ij} + \left(-i\dot{\bar{R}}_j\bar{P}_j + i\dot{\gamma}_j\right)\left(H - HS^{-1}S\right)_{ij} \\
&= \left(2\alpha\dot{\bar{R}}_i - i\dot{\bar{P}}_i\right)\left(H^{10} - S^{10}S^{-1}H\right)_{ij} + \left(2\alpha\dot{\bar{R}}_j + i\dot{\bar{P}}_j\right)\left(H^{01} - HS^{-1}S^{01}\right)_{ij}
\end{aligned} \tag{12}$$

and derive the condition of quantum energy conservation as follows

$$0 = \mathbf{C}^{\dagger}\left(\dot{\mathbf{H}} - \dot{\tilde{\mathbf{S}}}\mathbf{S}^{-1}\mathbf{H} - \mathbf{H}\mathbf{S}^{-1}\dot{\mathbf{S}}\right)\mathbf{C} = \sum_i^N \left(4\alpha\dot{\bar{R}}_i \cdot \mathrm{Re}(Z_i) + 2\dot{\bar{P}}_i \cdot \mathrm{Im}(Z_i)\right) \tag{13}$$

where $N$ is the number of TBFs on this single PES, and elements of vector $\mathbf{Z}$ is defined by

$$Z_i = \sum_{j=1}^N C_i^*\left(H_{ij}^{10} - \left(S^{10}S^{-1}H\right)_{ij}\right)C_j \tag{14}$$

For convenience, we have introduced moments matrix defined by

$$O_{ij}^{mn} \equiv \left\langle \chi_i(R;\bar{R}_i,\bar{P}_i,\gamma_i,\alpha) \middle| (R-\bar{R}_i)^m \cdot \hat{\mathbf{O}} \cdot (R-\bar{R}_j)^n \middle| \chi_j(R;\bar{R}_j,\bar{P}_j,\gamma_j,\alpha) \right\rangle_R \tag{15}$$

Eq. (13) is the primary result obtained in the paper. There are multiple solutions to (13). Here we consider one sufficient condition by assuming the identity in the summation $\sum_{i=1}^{N}$ holds for each $i$th TBF, i.e.

$$0 = \left(4\alpha \dot{\bar{R}}_i \cdot \text{Re}(Z_i) + 2\dot{\bar{P}}_i \cdot \text{Im}(Z_i)\right) \quad (16)$$

From condition (16), one can derive the EOM governing the propagation of $\dot{\bar{R}}_i$ and $\dot{\bar{P}}_i$ for each TBF,

$$\begin{aligned} \dot{\bar{R}}_i &= 2\lambda_i \cdot \text{Im}(Z_i) \\ \dot{\bar{P}}_i &= -4\alpha\lambda_i \cdot \text{Re}(Z_i) \end{aligned} \quad (17)$$

where $\lambda_i$ can be any arbitrary constant. Note that EOM (17) is a sufficient condition for (16) which guarantees the conservation of quantum mechanical energy.

For numerical convenience required by various codes and to demonstrate the quantum-classical mapping, we derive some classical approximations to the quantum EOM (17). In the simplest classical lime of $N=1$ (i.e. only one TBF is used in FMS dynamics), Eq. (17) simplifies to

$$\begin{aligned} \dot{\bar{R}} &= \frac{\bar{P}}{m} \\ \dot{\bar{P}} &= -\frac{d}{d\bar{R}}\left\langle \chi(R;\bar{R},\bar{P},\gamma)\middle|\mathbf{V}(R)\middle|\chi(R;\bar{R},\bar{P},\gamma)\right\rangle \end{aligned} \quad (18)$$

EOM (18) is very similar to the classical one except that potential $V(\bar{R}_i)$ in classical EOM (5) is replaced by its expectation value $\bar{V} \equiv \left\langle \chi(R;\bar{R},\bar{P},\gamma)\middle|\mathbf{V}(R)\middle|\chi(R;\bar{R},\bar{P},\gamma)\right\rangle$ weighted by the TBF $\chi$, which is similar to the role of effective potential as used in various semiclassical theories.[50] Interestingly, the identical form to (18) has been used in the literature for other important goal. For example, in CCS, (18) was used (which was referred to as averaged Hamiltonian with quantum correction) to successfully generate smooth potential to guide nuclear dynamics and replace Coulomb singularities by quardratic potential minima.[32]

We now consider another classical high-temperature limit by neglecting overlap between TBFs, i.e. $S_{ij} = 0$ for $i \neq j$, which approximates the dynamics when dimensionality of the system is high and all TBFs diverge very fast. One can show that the *moments* of overlap matrix in (16) becomes

$$S_{ij}^{10} = S_{ij} \cdot \left( -\frac{1}{2}(\bar{R}_i - \bar{R}_j) - \frac{i}{4\alpha}(\bar{P}_i - \bar{P}_j) \right) \quad (19)$$

which is 0 for both off-diagonal (because $S_{ij} = 0$) as well as diagonal matrix elements (because $\bar{R}_i = \bar{R}_j$). In this limit, one can easily show that EOM (17) becomes

$$\dot{\bar{R}}_i = |C_i|^2 \cdot \frac{\bar{P}_i}{m}$$
$$\dot{\bar{P}}_i = -|C_i|^2 \cdot \frac{dV_{ii}}{d\bar{R}_i} \quad (20)$$

where $V_{ii} \equiv \langle \chi_i(R; \bar{R}_i, \bar{P}_i, \gamma_i) | \mathbf{V}(R) | \chi_i(R; \bar{R}_i, \bar{P}_i, \gamma_i) \rangle$ as in (18). Actually, EOM (18) is a special case of (20) by setting population to 1 when $N=1$. Eq. (20) implies that the quantum force experienced by one TBF depends on the amplitude of all other TBFs. This is in contrary to the classical picture in which force should be independent on the population. However, quantum mechanically, all TBFs are entangled together such that amplitude does appear in the expression of force. Therefore, replacement of the classical force on the Gaussian center of TBF by the corresponding quantum mechanical expectation value introduces new kind of non-local force into our EOM.

In general, $S_{ij} \neq 0$ due to the coherence and interaction between neighboring TBFs. The quantum force becomes much more complicated as in (17). Briefly speaking, the conventional definition of force as spatial derivative of potential energy on single point is not enough and not complete. One should take into account two factors in computing the quantum force: non-local interaction of the PESs on the Gaussian centers of TBFs (e.g. use of expectation value $V_{ii}$ of PES), and coherence between neighboring TBFs (e.g. through the factor $\mathbf{S}^{-1}$).

*C. Generalization to ab initio potential and design of error function*

The existence of arbitrary constant $\lambda_i$ in the quantum EOM reflects the nature of non-unique mapping between quantum and classical dynamics. To uniquely determine the value of $\lambda_i$ as well as to better capture the classical limit using the current quantum EOM, we design and numericall implement a new error function

$$\sum_{i=1}^{N} \left( E_i^{cl}(t_2) - E_i^{cl}(t_1) \right)^2 \quad (21)$$

where $t_2 = t_1 + \Delta t$ is the increase in time for each propagation step of molecular dynamics. $E_i^{cl}$ is classical energy of each TBF $i$. Minimizing the error function (21) leads to a unique choice of $\lambda_i$, and thus quantum EOM, which minimizes the fluctuation in classical total energy.

Note that, the total classical energy is not conserved using limited number of TBF in FMS, because Wigner finite temperature distribution is used to generate initial conditions of bundle of TBFs. The way of generating multiple trajectories with classical energy obeying Wigner distribution can better describe quantum coherence, and has been used in many other

MQC methods. While population of each FMS TBF varies in the propagation, the total classical energy fluctuates too. Minimization of the error function (21) therefore reduces such fluctuation and retains the flavor of classical MD in addition to conserving quantum energy.

For simplicity, one-dimensional one-state potential model has been assumed in the above derivation. Generalization to arbitrary and/or ab initio potential models is straightforward. For multidimensional PES, position and momentum in Eq. (11) are vectors, and quantum EOM (17) is written by

$$\dot{\bar{R}}_{i\rho} = 2\lambda_i \cdot \text{Im}(Z_{i\rho}) \qquad (22)$$

where $\rho$ is the index of DOF. $Z_{i\rho}$ can be evaluated as follows

$$Z_{i\rho} = \sum_{j=1}^{N} C_i^* \left( H_{i\rho,j}^{10} - \sum_{m=1}^{N} \sum_{n=1}^{N} S_{i\rho,m}^{10} S_{m,n}^{-1} H_{n,j} \right) C_j$$

$$H_{i\rho,j}^{10} \equiv \left\langle \chi_i(R; \bar{R}_i, \bar{P}_i, \gamma_i, \alpha) \middle| (R_\rho - \bar{R}_{i\rho}) \mathbf{H} \middle| \chi_j(R; \bar{R}_j, \bar{P}_j, \gamma_j, \alpha) \right\rangle_R \qquad (23)$$

In the above expression, the multidimensional moments of Hamiltonian can be numerically evaluated by

$$\frac{\partial}{\partial \bar{R}_{i\rho}} H_{ij} = \left\langle \chi_i(R; \bar{R}_i, \bar{P}_i, \gamma_i, \alpha) \middle| \left( 2\alpha(R_\rho - \bar{R}_{i\rho}) + i\bar{P}_{i\rho} \right) \mathbf{H} \middle| \chi_j(R; \bar{R}_j, \bar{P}_j, \gamma_j, \alpha) \right\rangle_R$$
$$= 2\alpha H_{i\rho,j}^{10} + i\bar{P}_{i\rho} H_{ij} \qquad (24)$$

where the first-order derivatives of $H_{ij}$ can be numerically calculated.

For multi-electronic state systems, one has to solve the exact solution for (9). The analytic derivation appears very forbidding, and numerical evaluation is expected to be much more expensive. We introduce another approximation by neglecting the interaction between different PESs. Equations of motion (17) are unchanged, only with the matrix element of $Z$ re-written in the form

$$Z_i = \sum_{\substack{j=1 \\ S'_j = S_i}}^{N} C_i^* \left( H_{ij}^{10} - \left( S^{10} S^{-1} H \right)_{ij} \right) C_j \qquad (25)$$

where $\sum_{j=1}^{N}$ is over all the TBFs on the same PES as $i$-th trajectory. The quantum energy is approximately conserved, better than using classical EOM, when nonadiabatic coupling is small. In the neighborhood of conical intersection, one expects conservation of quantum energy to be less achieved. The procedure can be improved by adding the amount of fluctuation of quantum energy into the error function in determining $\lambda_i$ variationally. To further reduce the computational cost, fluctuation in the total quantum energy using classical EOM is calculated for each propagation time step. Quantum EOM is only used when change in the classical energy exceeds certain threshold value $\delta$. Therefore, when the system is in the classical allowed region, fluctuation in classical energy is usually small, and quantum EOM is used much less frequently.

### III. MODELS AND DISCUSSIONS

We choose the one-dimensional double well potential to test the new quantum EOM (17) and to compare it with the classical Hamilton's EOM as well as analytical solutions. The double well potential is given by

$$V(R) = V_0 + D(R - R_0)^4 - C(R - R_0)^2. \qquad (26)$$

where the values of parameters $C$ and $D$ are chosen corresponding to the potential for hydrogen atom transfer in malonaldehyde when coupled to harmonic oscillators.[74] This potential model is a benchmark showcase for a number of important model tests, and is of particular interest and importance to the study of tunneling effects.

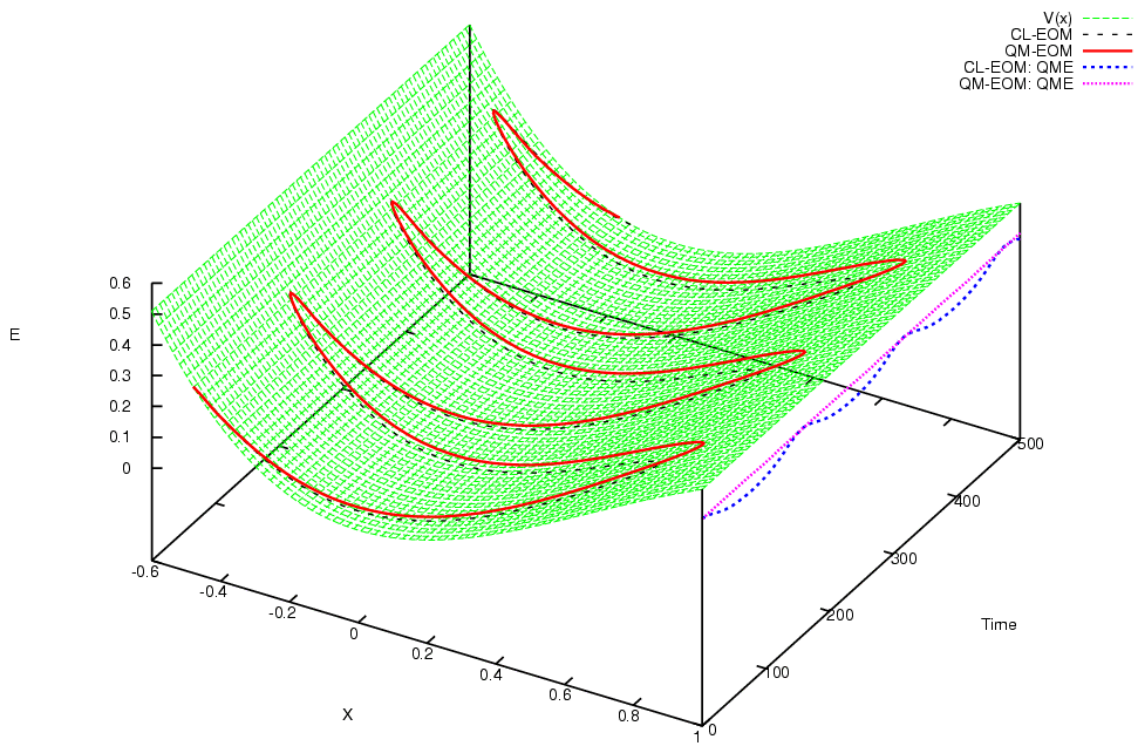

**Figure 1:** Phase space dynamics and total energy with one basis function for double well potential. Solid red line shows the position *x(t)* of the Gaussian center of TBF obeying quantum EOM, and black line corresponds to classical EOM. Both lines are plotted and embedded into the PES for clearer observation. Pink dotted line shows the total quantum energy using quantum EOM, while the blue dotted lines corresponds to the use of classical EOM.

In Figure 1, dynamics of phase space variable *R(t)* is plotted. Comparison is made between the trajectory *R(t)* using quantum and classical EOM. For clearer observation , plots of trajectory $R(t)$ is embedded into the PES of the double well. As shown by the plots, quantum EOM presented in this paper is close to the classical trajectory in the classical region. The essential difference between the two types of EOM is clear when the particle approaches the classical turning point near $x$=0.8. Red line penetrates deeper into the potential barrier, which represents the tunneling behavior in the case of low kinetic energy. The deeper penetration of the particle is due to the use of averaged potential in (18), where replacement of potential of the Gaussian center by the corresponding expectation or averaged potential leads to smoothing effect. The similar observation has been utilized in the earlier paper by Shalashilin.[75] In this paper, it is shown that such a smoothed potential is actually a first order classical approximation to the quantum EOM. In the case of single TBF, such a first order approximation becomes exact and leads to the strict conservation of quantum energy as well as propagation with semiclassical nature. In Figure 1, total quantum mechanical energy $E_{QM}(t)$ is also plotted. It is clearly shown that total quantum energy is only conserved when new quantum EOM is used to propagate the Gaussian centers $(x,p)$.

In Figure 2, two initial TBFs are used for double well potential model. On the left are shown the plots of quantum energy. The FMS results using classical EOM are plotted in red (for total energy) and purple (for each individual TBF), and those using quantum EOM in green (for total energy) and blue (for each TBF). The threshold value $\delta$ is set to 0.001, such that classical EOM is used in place of quantum one if the change in the classical energy for each time step does not exceed d. Such a numerical approximation still guarantees that the new quantum EOM conserves quantum energy much better than the classical EOM does. Classical EOM conserves the classical energy of each individual TBF, but does not conserve quantum counterpart. In the subplot on the right, population of each TBF is plotted. Quantum

EOM usually features less frequent population transfer among various TBFs, which shares some similarity with the fewest switch version of surface hopping.[76] Though the derivations of two schemes are independent and physical pictures are very different, this similarity reveals one general physical intuition of dynamics: avoiding unnecessary population transfer has the advantage of reducing quantum oscillation and thus improves the simulation results.

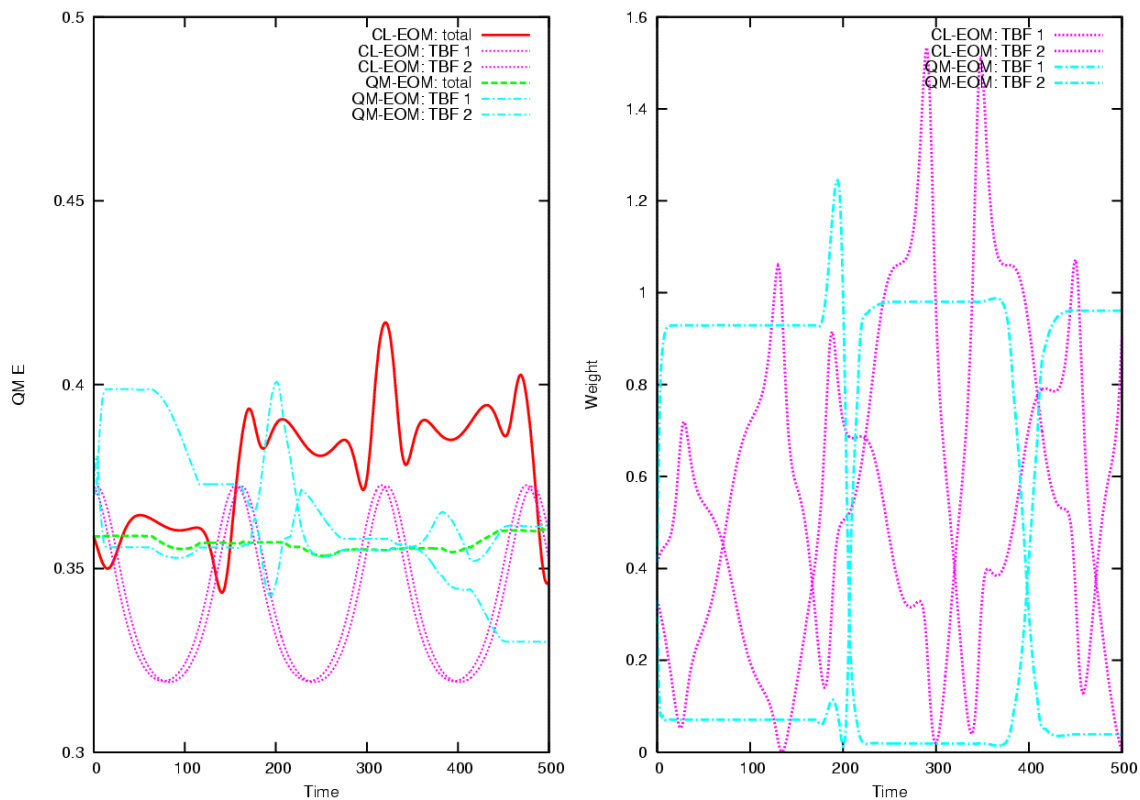

**Figure 2:** Energy and position of Gaussian centers of TBFs using quantum and classical EOM. Two initial TBFs are used. On the left is shown the plots of quantum energy of total system and of each individual TBF. The FMS results using classical EOM are plotted in red (for total energy) and purple (for each individual TBF), and those using quantum EOM in green (for total energy) and blue (for each TBF). The threshold value $\delta$ is set to 0.01, such that classical EOM is used in place of quantum one if the change in the classical energy for each time step does not exceed $\delta$. In the right subplot, weight of each TBF is plotted using

quantum EOM (blue) and classical EOM (purple) respectively. Quantum EOM usually features less frequent population transfer among various TBFs.

We now examine the performance of the quantum EOM with a two-dimensional, two-state conical intersection model. The model was originally introduced by Ferretti, *et al.*[77-82] to describe a collinear triatomic *ABA*. It provides a useful testing bed for study and comparison of nonadiabatic simulation schemes. There are two diabatic electronic states, henceforth referred to as |1> and |2>, and two coordinates, *X* and *Y*, corresponding to symmetric and antisymmetric stretching. The diabatic potential matrix elements are given by

$$V_{11}(X,Y) = \frac{1}{2}k_x(X-X_1)^2 + \frac{1}{2}k_y Y^2$$
$$V_{22}(X,Y) = \frac{1}{2}k_x(X-X_2)^2 + \frac{1}{2}k_y Y^2 + \Delta \quad (27)$$
$$V_{12}(X,Y) = V_{21}(X,Y) = \gamma Y e^{-\alpha(X-X_3)^2 - \beta Y^2}$$

The interstate coupling is controlled by the parameter $\gamma$. The parameters in the Hamiltonian are chosen same as in the previous work,[83,84] i.e. we choose $k_x$=0.01, $k_y$=0.1, $\Delta$=0.01, $\alpha$=3, $\beta$=1.5, $\gamma$=0.01, and $X_3$=3. The initial wavepacket is associated with the diabatic state |1> and the total simulation time corresponds roughly to one half-period along the *X* direction. The width of the initial Gaussian wavepacket is chosen to be 22.2 and 12.9 bohr$^{-2}$ along *X* and *Y* directions, respectively. The initial trajectory basis functions are sampled randomly from the Wigner corresponding to the desired initial wavefunction and the TBF widths are taken to be the same as those of the initial wavepacket.

In Figure 3, comparison of classical and new quantum EOMs are compared using the Persico model (27). For this specific model, since both diabatic states are harmonic, FMS employing classical EOM for the Gaussian centers of each TBF conserves total quantum mechanical energy, as shown in the left subplot of Figure 3. Using quantum EOM does not break this conservation. The more interesting point is shown in the right subplot in which plots of classical energy using quantum EOM (red) and classical EOM (black) are compared. Because of the use of error function (21), classical energy is better conserved as a bonus point of the new quantum EOM.

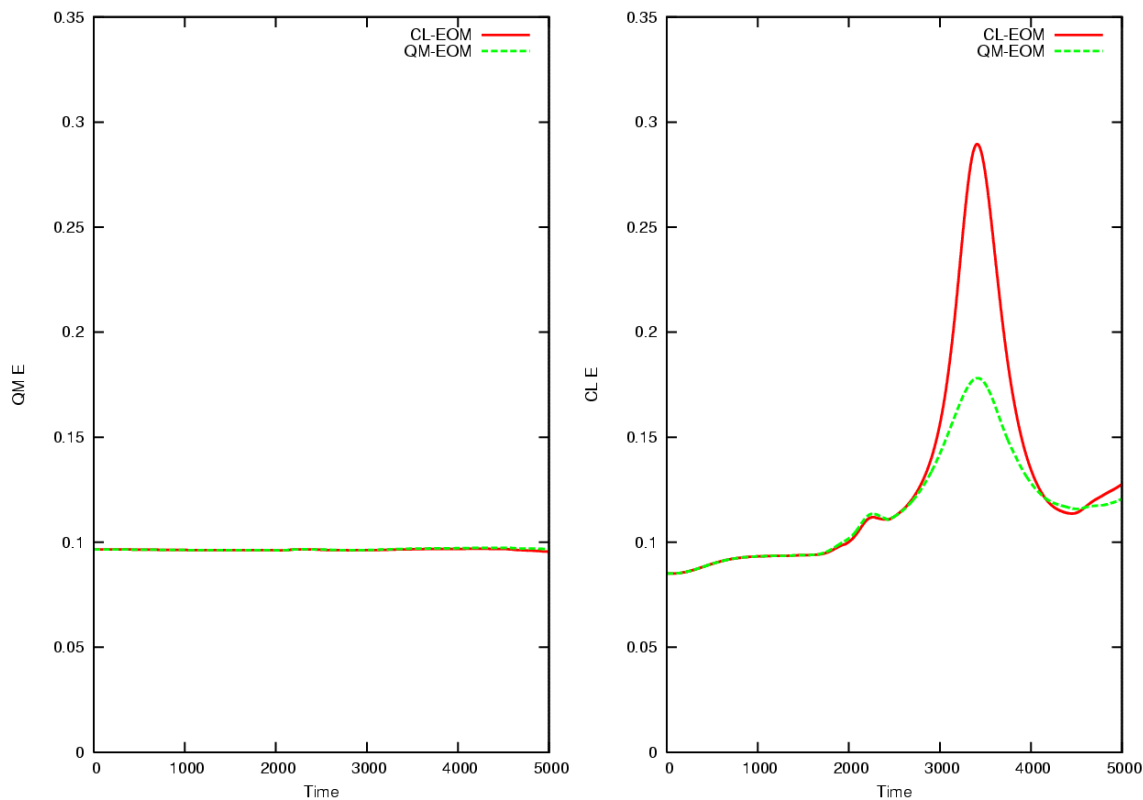

**Figure 3:** Calculation of quantum and classical energies are shown for 2-state 2-dimension Persico model with strong nonadiabatic coupling. Subplots on the left and right show the quantum and classical energy of the system respectively. In each subplot, classical and quantum EOM are compared. Since the two diabatic states are harmonic, quantum energy is conserved using either type of EOM in FMS. The fluctuation of classical energy using quantum EOM is much smaller because of the error function employed in the new method.

## IV. CONCLUSIONS

Developing new quantum wavepacket dynamics capable of describing nonadiabatic transitions and quantum tunneling effects has been known as a major challenge. Conservation of quantum mechanical energy is difficult when one uses nuclear trajectories with classical equations of motion. In this paper, we presented a new set of equations of motion which conserve the quantum energy of the trajectories in the molecular dynamics simulation. The new set of equations is implemented in FMS method to propagate the Gaussian center of the nuclear basis functions. One-dimensional double well potential model has been used to demonstrate the conservation of quantum mechanical energy and other characteristics of the new equations. We employ the error function minimization technique to choose the optimal classical-like quantum propagation (i.e. determining the value of λ in (17)). Numerical simulations on the two-state two-dimensional potential model with strong nonadiabatic coupling shows that the error function minimization technique minimizes the fluctuation in the classical energy.

The new quantum EOM plays important roles in describing quantum tunneling. Including tunneling in a classical trajectory simulation of the full MD simulation has attracted much attention.[27,85,86] In the earlier work by Makri and Miller,[76] it is assumed that each classical trajectory that oscillates in the potential well tunnels out with probability $e^{-2\theta}$ where θ is the classical action integral through the potential barrier. In our simulation, FMS is used such that each TBF is assigned with exact quantum amplitude. Additionally, implementing the new quantum EOM enables each TBF propagate quantum mechanically near classical turning points as shown in Figure 1. The global quantum force experienced by each TBF promotes energy redistribution among the bundle of trajectories. Neighboring trajectories thus provide extra energy to the specific trajectory to help it tunnel through the potential barrier higher than the energy of the single trajectory itself. In the work by Makri and Miller,[76] a tunneling model was presented in which the classical trajectory evolving in classically allowed region has a probability for making instantaneous transition to another classically allowed region separated by a potential barrier. In our model, exact quantum wavepacket is propagated with the Gaussian centers of each basis function obeying the rigorous quantum EOM.

The new quantum EOM provides a better way to treat the self-consistency between the quantum and classical DOF in MQC. Each quantum mechanical DOF corresponds to an infinitely dimensional Hilbert space. Quantum-classical correspondence deals with the mapping of Hilbert space onto a phase space. Various MQC approaches have been explored to avoid the high computational cost of exact quantum solutions. Separating system into classical and quantum parts is usually validated by the large mass ratio or relative energies associated with different DOF. Various ways of coupling between quantum and classical DOF lead to different MQC schemes like mean-field and surface hopping methods. In spite of huge success and convenience gained, MQC suffers from several problems, one of which is back reaction of classical on the quantum DOF. Our new EOM incorporates quantum force into classical-like EOM while obeying self-consistent back reaction between quantum and classical DOF. Treating various DOF on the same footing and with correct quantum limit is important for problems like proton motion between one water molecules to the next involving tunneling. The new set of EOM differs from Bohmian dynamics in that the former one incorporates nonlocal quantum mechanical force in a more practical and robust way. The quantum force $-4\alpha\lambda_i \cdot \text{Re}(Z_i)$ in (17) determines the time derivative of momentum of the centroid of each TBF. The nonlocal feature is taken accounted into the quantum force when the individual population $c_i$ of and overlapping among all the TBFs are present in the EOM. Note that the calculation of such quantum force is very affordable because the overlapping matrix is already calculated and stored in FMS dynamics. This is consistent with the feature of FMS which approaches exact quantum dynamics when basis set is increased and discrete trajectory-evolution becomes exact quantum fluid propagation as in Bohmian dynamics.